\documentclass[11pt]{article}
\usepackage{amsmath}
\usepackage{amsfonts}
\usepackage{amssymb}
\usepackage{epsfig}
\usepackage{times}

\newcommand{\ol}{\overline}

\renewcommand{\l}{\newline\null}
\def\figskip{\vskip .5cm plus 3mm minus 2mm}
\def\hbar{\not{\hbox{\kern-2.3pt $h$}}}
\def\psl{\not{\hbox{\kern-2.3pt $p$}}}
\def\Psl{\not{\hbox{\kern-2.3pt $P$}}}
\def\ksl{\not{\hbox{\kern-2.3pt $k$}}}
\def\qsl{\not{\hbox{\kern-2.3pt $q$}}}
\newcounter{hours}\newcounter{minutes}

\begin{document}
%
\begin{titlepage}
July 2002 \hfill PAR-LPTHE 02/35
%
%
\vskip 4cm
{\baselineskip 17pt
\centerline{
{\bf NEUTRINO MIXING ANGLES AND EIGENSTATES;}}
\centerline{{\bf $\mathbf{CP}$ PROPERTIES AND MASS HIERARCHIES}}
}
\vskip .5cm
\centerline{B. Machet
     \footnote[1]{Member of `Centre National de la Recherche Scientifique'}
     \footnote[2]{E-mail: machet@lpthe.jussieu.fr}
     }
\vskip 5mm
\centerline{{\em Laboratoire de Physique Th\'eorique et Hautes \'Energies}
     \footnote[3]{LPTHE tour 16\,/\,1$^{er}\!$ \'etage,
          Universit\'e P. et M. Curie, BP 126, 4 place Jussieu,
          F-75252 Paris Cedex 05 (France)}
}
\centerline{\em Universit\'es Pierre et Marie Curie (Paris 6) et Denis
Diderot (Paris 7)}
\centerline{\em Unit\'e associ\'ee au CNRS UMR 7589}
\vskip 1.2cm
{\bf Abstract:}
In the presence of independent generations of leptons, I show that the same
type of ambiguity in the mass spectrum arises as was discussed in
\cite{Machet1} for neutral kaons.
It results from the freedom to add to their Majorana
mass matrix, usually taken to be symmetric, an antisymmetric term  which
vanishes as soon as fermions belonging to different generations
anticommute.

In the simple examples proposed, dealing with two generations,
this procedure introduces an extra (mass) parameter $\rho$, which
is shown to connect the ($CP$ violating) mixing angle to the hierarchy of
neutrino masses.
We use this opportunity to investigate the relations between the two; in
particular, large hierarchies are no longer preferentially
attached to small mixing angles; this can be relevant for the
``Large Mixing Angle'' solution strongly advocated by recent
experiments on neutrinos oscillations \cite{LMA}.

I discuss how the $\rho$ parameter could be fixed, which appears,
in the absence of a substructure for leptons, still more delicate
than for kaons.

\smallskip

\vfill
\null\hfil\epsffile{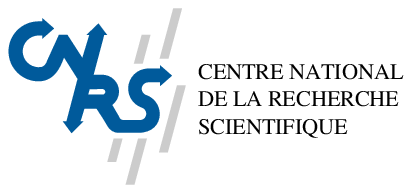}
\end{titlepage}
%
%
%

\section{Introduction}
\label{section:introduction}
%
It was shown in \cite{Machet1} that the spectrum of the pair of
charge-conjugate electrically neutral kaons
\footnote{and it was extended there, too, to Higgs-like doublets.}
 can be ambiguous as
soon as they are considered to be independent and commuting
fields. The introduction, in the Lagrangian, of a formally
vanishing ``mass term''  proportional to their commutator induces
a  degeneracy for the set of basis in which the kinetic and mass
terms become simultaneously diagonalizable.

A $U(1)$ group was pointed out, the phase  $\theta$ of which
became, by the process mentioned above,  the  indirect
$CP$-violating parameter of the neutral kaon system. Mass
eigenstates can vary from  self- (or anti-self-) conjugate states
to states of definite ``flavor number'' ($K^0$ and $\ol{K^0}$) and
the hierarchical pattern of masses becomes a function of $\theta$.

The different possible behaviors of mass eigenstates by charge
conjugation suggested in \cite{Machet1} a parallel with Majorana
(self-, or anti-self-conjugate) and / or Weyl / Dirac neutrinos.
The purpose of this letter is to show, with  very simple example in the case
of two generations, how, indeed, the same type of ambiguity
arises for neutrinos, and how the departure from Majorana states is
controlled by a phase which determines their hierarchy pattern and $CP$
properties.

\section{The mass matrix for neutrinos}
\label{section:mmat}

\subsection{The customary framework \cite{Bilenky}}
\label{subsection:custom}

The Majorana mass matrix for neutrinos is usually taken to be symmetric.

Let us consider the example of a pure Majorana mass terms for two
generations of neutrinos. We accordingly introduce the vector
of left-handed neutrinos \footnote{The indices have been named
here $e$ and $\mu$ only to recall that they are {\em flavor}
indices; they do not relate specifically to the electron or to the
muon neutrinos.}
\begin{equation}
n_L = \left( \begin{array}{c} \nu_{eL} \cr
                              \nu_{{\mu} L} \end{array}\right)
\end{equation}
and the mass term (the superscript ``$c$'' means {\em ``charge
conjugate''})
\begin{equation}
{\cal L}^m_0 = -\frac{1}{2}\; \ol{(n_L)^c}\; M_0\; n_L + h.c.;
\label{eq:Lm0}
\end{equation}
$M_0$ is the $2 \times 2$ mass matrix
\begin{equation}
M_0 = \left( \begin{array}{cc} m_1 & d \cr
                                d  & m_2 \end{array} \right)
\label{eq:M0}
\end{equation}
that we shall take real for the sake of simplicity.
(\ref{eq:Lm0}) violates, as usual, the conservation of the
leptonic number.

That $M_0$ is taken to be symmetric results from the anticommutation
relations between fermions \cite{Bilenky}
\begin{equation}
\ol{(\nu_{eL})^c} \nu_{\mu L} - \ol{(\nu_{\mu L})^c} \nu_{eL} = 0.
\label{eq:anticommute}
\end{equation}
Consequently, $M_0$, can be diagonalised  by a unitary matrix
$V^0$ according to
\begin{equation}
V^{0T} M_0 V^0 = D^0 = diag(M^0_1, M^0_2),
\end{equation}
where the superscript ``$T$'' means {\em ``transposed''}; $V^0$ is
given by:
\begin{equation}
V^0 = \left( \begin{array}{rr} c_\theta &  -s_\theta \cr
                               s_\theta & c_\theta
                               \end{array}\right),
\label{eq:V0}
\end{equation}
with
\begin{equation}
\tan(2\theta) = \frac{2d}{m_1 - m_2}.
\end{equation}
${\cal L}^0_m$ given in (\ref{eq:Lm0}) rewrites
\begin{equation}
{\cal L}^0_m = -\frac{1}{2}\; \ol{N^0}\; D^0\; N^0 + h.c.,
\end{equation}
in terms of  the mass eigenstates $N^0$, which are Majorana
neutrinos ($N^0_L = (N^0_R)^c$), given by \cite{Bilenky}
\begin{equation}
N^0 =  V^{0\dagger} n_L + (V^{0\dagger} n_L)^c.
\label{eq:majo1}
\end{equation}
$\theta$ is the ``mixing angle'' for neutrinos \cite{Bilenky}.

The eigenmasses $M_1$ and $M_2$ are given, in terms of $\theta$,
by
\begin{eqnarray}
M_1 &=& \left\vert\frac{m_1\cos^2\theta -
              m_2\sin^2\theta}{\cos^2\theta-\sin^2\theta}\right\vert,\cr
M_2 &=& \left\vert\frac{m_2\cos^2\theta -
                  m_1\sin^2\theta}{\cos^2\theta-\sin^2\theta}\right\vert;
\label{eq:M10M20}
\end{eqnarray}
we have used the fact that both eigenvalues can always be made
positive by eventually multiplying $V^0$ by $diag(1,i)$ or
$diag(i,1)$.

The spectrum ($M_1$ and $M_2$) is drawn on Fig.~1 as a function of
$\theta$, for $m_1=2$ and $m_2=4$. On Fig.~2a and 2b are drawn
respectively the ratios $M_1/M_2$ and $M_2/M_1$.

One of the eigenmasses vanishes for $d^2 = m_1m_2$; the other is
then given by $(m_1 + m_2)$.

The case of maximal mixing ($\vert\theta\vert = \pi/4$)
corresponds either to $d \rightarrow \infty$ or to $m_2=m_1=m,
\forall d$; in the first case, the eigenmasses are $\vert\pm
d\vert \rightarrow \infty$ in the last case they are $\vert m \pm
d \vert$
\footnote{The case $m_1 = m_2 = m$ is special since, then,
as can be seen directly, the mixing angle $\theta$ is fixed to its
maximal value $\pi/4$.}
.
\figskip \vbox{
\begin{center}
\epsfig{file=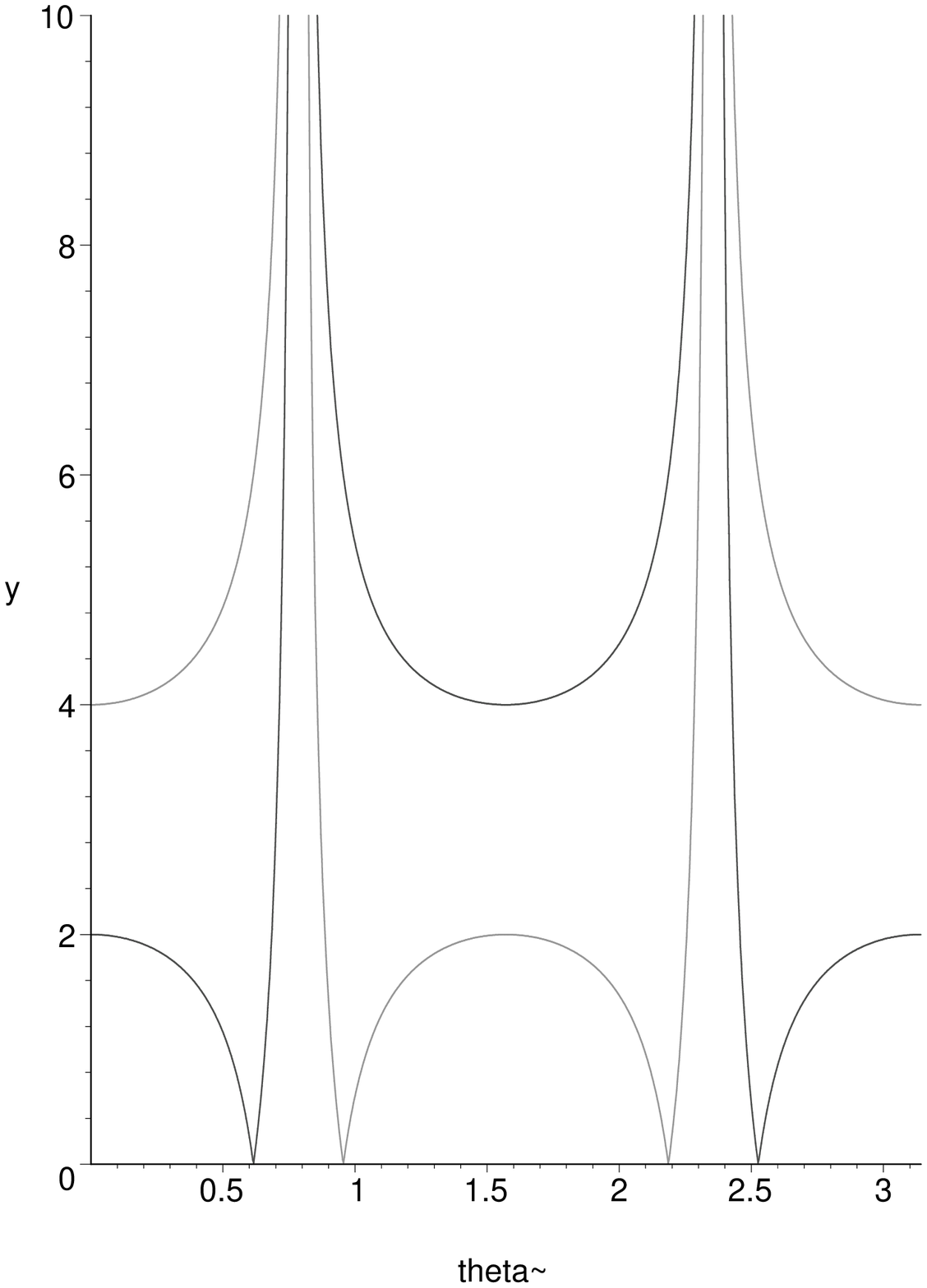,height=7truecm,width=8truecm}
\figskip
{\em Fig.~1:$M_1$ and $M_2$ as functions of $\theta$, for $m_1=2$,
$m_2=4$ and $\rho=0$}
\end{center}
}
\figskip
\vbox{
\begin{center}
\epsfig{file=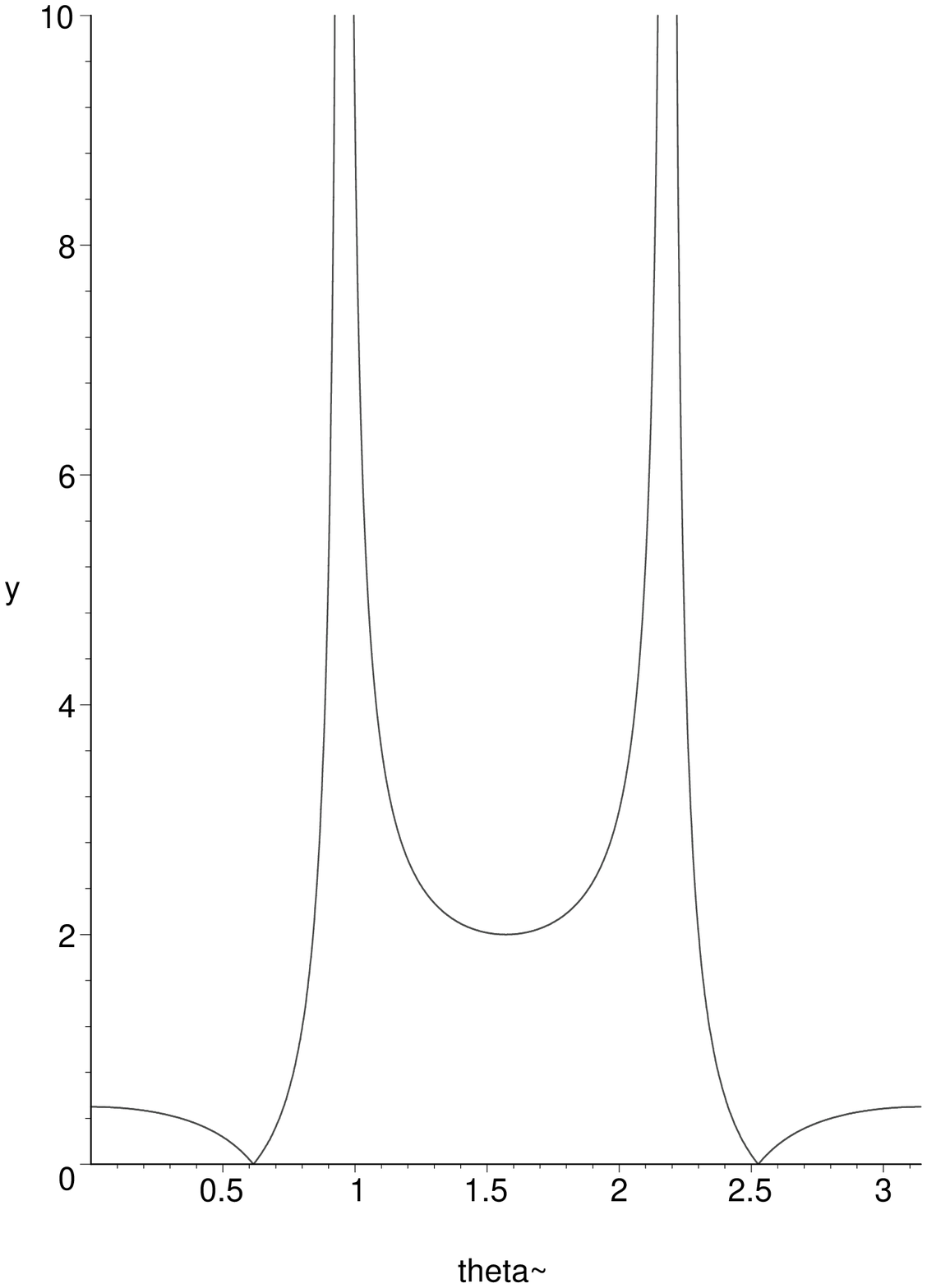,height=7truecm,width=8truecm}
\figskip
{\em Fig.~2a:$M_1/ M_2$ as function of $\theta$ for $m_1=2$, $m_2=4$ and
$\rho=0$}
\end{center}
}
\figskip
\vbox{
\begin{center}
\epsfig{file=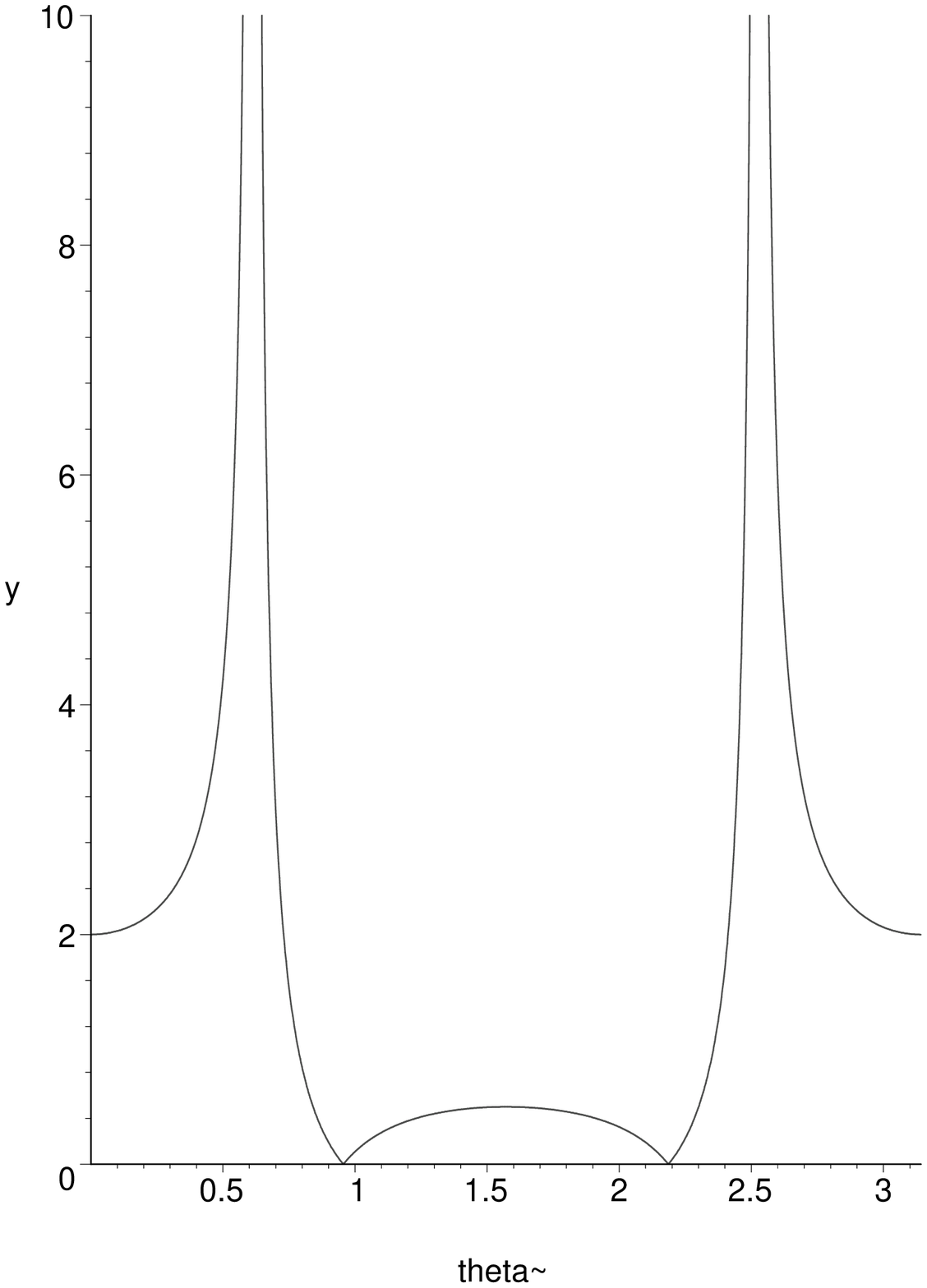,height=7truecm,width=8truecm}
\figskip
{\em Fig.~2b:$M_2/M_1$ as function of $\theta$ for $m_1=2$, $m_2=4$ and
$\rho=0$}
\end{center}
}
%
\subsection{Independent generations of fermions}
\label{subsection:indep}

Taking for granted that, the two generations of fermions being supposed to be
independent, the corresponding fields anticommute,  the  relation
(\ref{eq:anticommute}) holds.
Instead of using, like in subsection \ref{subsection:custom} this property
to take {\em a priori} the Majorana mass
matrix to be symmetric, we shall use below the freedom that it yields
to add to the latter a vanishing antisymmetric term.

Accordingly, the two mass matrices given respectively by $M_0$ in
(\ref{eq:M0}) and by $M$ in (\ref{eq:Mgen})  below
\begin{equation}
M = \left( \begin{array}{cc} m_1 & d+\rho \cr
                            d-\rho & m_2  \end{array} \right),
\label{eq:Mgen}
\end{equation}
in which, for simplification, we also take $\rho$ to be real,
only differ by what corresponds to a vanishing mass term in the Lagrangian
\footnote{It also corresponds to the simplest type of ``horizontal
interaction'', and minimal in the sense that it vanishes.}
\begin{equation}
-\rho/2\left(\ol{(\nu_{1L})^c} \nu_{2L} - \ol{(\nu_{2L})^c}
\nu_{1L}\right).
\label{eq:vmt}
\end{equation}
By this procedure, the mass Lagrangian goes from ${\cal L}^0_m$ given by
(\ref{eq:Lm0}) to
${\cal L}_m$ given below
\begin{equation}
{\cal L}_m = -\frac{1}{2}\; \ol{(n_L)^c}\; M\; n_L + h.c..
\label{eq:M}
\end{equation}
%

\subsection{Diagonalization}
\label{subsection:diag}
\subsubsection{Bi-unitary transformation}
\label{subsubsec:biunit}

The mass matrix $M$ in (\ref{eq:Mgen}), which is no longer symmetric
(neither hermitian), cannot be diagonalised any more by a single
unitary transformation. It is however mandatory that the
diagonalization be performed with unitary matrices in order that
the kinetic terms  stay diagonal \footnote{One recalls that the
kinetic terms satisfy $i\, \ol{(\nu_L)^c} \gamma^\mu \partial_\mu
(\nu_L)^c = i\, \ol{\nu_L} \gamma^\mu
\partial_\mu \nu_L$. \label{foot:kin}}.
One then accordingly proceeds with a bi-unitary transformation
\cite{Mohapatra}.

There exist two unitary matrices $U$ and $V$ such that
\begin{equation}
U^\dagger M V = D = diag(M_1,M_2).
\label{eq:bidiag}
\end{equation}
$U$ and $V$ diagonalize respectively the hermitian matrices $M
M^\dagger$ and $M^\dagger M$.

${\cal L}_m$ rewrites
\begin{equation}
{\cal L}_m = -\frac{1}{2}\; \ol{(U^T n_L)^c}\; D\; (V^\dagger n_L)
+ h.c..
\end{equation}
This leads to the two changes of basis below, which yield
left-handed and right-handed mass eigenstates, according to
\begin{eqnarray}
N_L &=& V^\dagger n_L, \cr
N_R &=& (U^T n_L)^c =  (U^T)^\dagger (n_L)^c = U^\ast (n_L)^c.
\label{eq:masseig1}
\end{eqnarray}
The kinetic terms stay diagonal (see footnote \ref{foot:kin})
since
\begin{equation}
\frac{i}{2}\,\left(\ol{N_L} \gamma^\mu\partial_\mu N_L +
     \ol{N_R} \gamma^\mu\partial_\mu N_R     \right)
           = i\, \ol{n_L} \gamma^\mu\partial_\mu n_L.
\end{equation}
The diagonalised mass matrix no longer connects Majorana neutrinos
but the Weyl spinors $N_R$ and $N_L$ ($N_L \not= (N_R)^c$).

The mixing angle for the left-handed neutrinos is given by the matrix $V$
and has {\em a priori} no  relation with $\theta$ defined by $V^0$ in
(\ref{eq:V0}).

The mass eigenvalues are no longer the solutions of the characteristic
equation of $M$ in (\ref{eq:Mgen}) but the ``square roots'' of
the eigenvalues of $M M^\dagger$.
They can be very different from the eigenvalues of $M_0$ in (\ref{eq:M0}).

Note that, unlike in (\ref{eq:majo1}), the Dirac spinor $N_L + N_R$ (see
(\ref{eq:masseig1})) cannot be taken as the set of mass eigenstates since
\begin{equation}
(\ol{N_L} + \ol{N_R}) D (N_L + N_R) =  \ol{(n_L)^c} M n_L +
\ol{n_L} V D U^\dagger (n_L)^c,
\label{eq:eqm1}
\end{equation}
the last contribution of which  is unsuitable.

\subsubsection{A short comment on oscillations}
\label{subsubsec:oscill}

In the $(N_L, N_R)$ basis, the Lagrangian (kinetic + mass terms) writes
(recall that $N_L$ and $N_R$ are themselves $2$-vectors)
\begin{equation}
{\cal L}= \frac{1}{2}\left(\begin{array}{cc} \ol{N_L} & \ol{N_R}
                                                       \end{array}\right)
\left(\begin{array}{cccc}  p\!\!/ & & -M_1 &  \cr
                              & p\!\!/ & & -M_2 \cr
                          -M_1 & & p\!\!/ & \cr
                              & -M_2 &  & p\!\!/ \end{array}\right)
\left(\begin{array}{c}   N_L \cr
                       N_R        \end{array}\right);
\end{equation}
while the kinetic terms are diagonal, the mass terms, which can only connect
``left'' to ``right'' Weyl fermions are now placed on the antidiagonal
blocks.

One cannot use any more the standard form of the Dirac propagators for
fermions; however, suppose that a $N_{1L}$ is created with energy momentum
$p$; its propagator, expressed by the series depicted in Fig.~3, writes
\begin{equation}
\frac {i}{p\!\!/} + \frac {i}{p\!\!/} iM_1 \frac {i}{p\!\!/} iM_1 \frac
{i}{p\!\!/} + \cdots = \frac{ip\!\!/}{p^2 -M_1^2}
\end{equation}
showing as expected a pole at $p^2 = M_1^2$. The left-handed Weyl fermion
$N_{1L}$ does propagate like a particle of mass $M_1$. A similar remark
applies to $N_{2L}$.
\figskip
\vbox{
\begin{center}
\epsfig{file=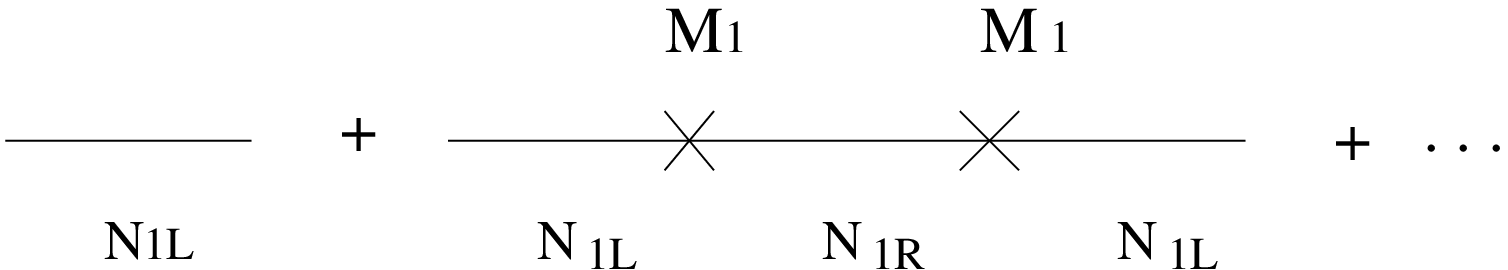,height=2truecm,width=8truecm}
\figskip
{\em Fig.~3: propagating the Weyl fermion $N_{1L}$.}
\end{center}
}
The phenomenon of oscillations proceeds as in the standard case:
if a neutrino of a
definite flavor is created at a space-time point $x_0$, its two Weyl
components, obtained through the simple inversion of (\ref{eq:masseig1}),
propagate
with different frequencies and wavelengths corresponding respectively to the
eigenmasses $M_1$ and $M_2$; after a certain time and distance of propagation,
at point $x$, a left-handed neutrino, for example, if detected, will not
have the same proportion of $N_{1L}$ and $N_{2L}$ as it had at point $x_0$.

\section{Examples}
\label{section:example}

We shall successively treat three examples of increasing complexity
corresponding to a mass matrix (\ref{eq:Mgen}) with $\rho \not = 0$: \l
- the case $m_2 = m_1, d \not= 0$;\l
- the case $m_2 \not= m_1, d=0$;\l
- the more general case $m_2 \not= m_1, d \not= 0$.

\subsection{The case $\mathbf{m_2 = m_1, d \not= 0}$}
\label{subsec:degen}

This very simple case, easily solvable, nevertheless exhibits most
of the properties that we want to emphasize: departure from the
status of Majorana neutrino and indirect $CP$ violation,
dependence of the spectrum on the mixing angle and, in this
precise case, one-to-one relationship between the latter and the
hierarchy of masses.

The mass matrix is
\begin{equation}
{\cal M} = \left( \begin{array}{cc} m   &  d+\rho  \cr
                            d-\rho &   m  \end{array} \right),
\label{eq:M1}
\end{equation}
and we choose the ``diagonalizing'' unitary matrices $U$ and $V$
to span, for conveniency, the two sets described by
\begin{equation}
A(\varphi) = \left( \begin{array}{rr}  c_\varphi  &  -s_\varphi   \cr
                                    s_\varphi  &  c_\varphi \end{array}\right),
\quad
B(\omega) = \left( \begin{array}{rr}  -c_\omega  &  s_\omega   \cr
                                    s_\omega  &  c_\omega
                                    \end{array}\right),
\label{eq:AB}
\end{equation}
respectively with determinants $+1$ and $-1$.

$\varphi$ is the mixing angle for leptons \cite{Bilenky}

%
\subsubsection{Case $\mathbf{d > m}$}
\label{subsubsec:dsupm}

The diagonalization equation
\begin{equation}
B^\dagger(\omega)\; {\cal M}\; A(\varphi) = diag(M_1, M_2)
\label{eq:diag1}
\end{equation}
is satisfied for
\begin{eqnarray}
&&\omega = \varphi + \pi/2, \cr && \tan(2\varphi) =
-\frac{m}{\rho}; \label{eq:rho1}
\end{eqnarray}
the mass eigenvalues are then
\begin{eqnarray}
M_1 &=& d + \frac{m}{\sin(2\varphi)}\cr
M_2 &=& d - \frac{m}{\sin(2\varphi)}.
\label{eq:mass1}
\end{eqnarray}
The condition $M_1 \geq 0$ requires $\sin(2\varphi) \leq -m/d$ or
$\sin(2\varphi) \geq 0$.
The condition $M_2 \geq 0$ requires $\sin(2\varphi) \geq m/d$ or
$\sin(2\varphi) \leq 0$.

Inside the interval $-m/d \leq \sin(2\varphi) \leq m/d$, one among
the two eigenmasses is negative; this problem is easily taken care
of: for $-m/d \leq \sin(2\varphi) \leq 0$ one goes to a positive
$M_1$ ($M_2$ is positive) by multiplying $A$ and $B$ by the matrix
$diag(i,1)$; for $0 \leq \sin(2\varphi) \leq m/d$, one goes to a
positive $M_2$ ($M_1$ is positive) by multiplying $A$ and $B$ by
the matrix $diag(1,i)$.

The spectrum is drawn on Fig.~4 for $m=2$ and $d=4$. Its
periodicity with respect to $\varphi$ has been used to draw the
picture for $\varphi \in [0,\pi/2]$ rather than $\varphi \in
[-\pi/4,\pi/4]$. It has been divided into three zones:\l - in the
central one, $\varphi \in [(1/2)\arcsin(m/d), \pi/2 -
(1/2)\arcsin(m/d)]$, $M_1 + M_2 = 2d$ and $|M_2 - M_1| = \vert
2m/\sin(2\varphi)\vert$. The extrema for the masses (one maximum
and one minimum) are respectively  $d+m$ and $d-m$ and correspond
to $\rho=0, \sin(2\varphi)=1$; the mass matrix $\cal M$ is then
symmetric and the mass eigenstates can also be taken as Majorana
neutrinos (see section \ref{section:WM});\l - in the two lateral
zones, $|M_2 - M_1| = 2d$ and $M_1 + M_2 =
  2m/\sin(2\varphi)$;\l
- the mass of one of the  neutrinos vanishes at the boundaries between
  the two zones above, which corresponds to $\rho^2 = d^2 - m^2$ or
 $\sin(2\varphi) = m/d$. The other mass is then equal to $2d$.
\figskip
\vbox{
\begin{center}
\epsfig{file=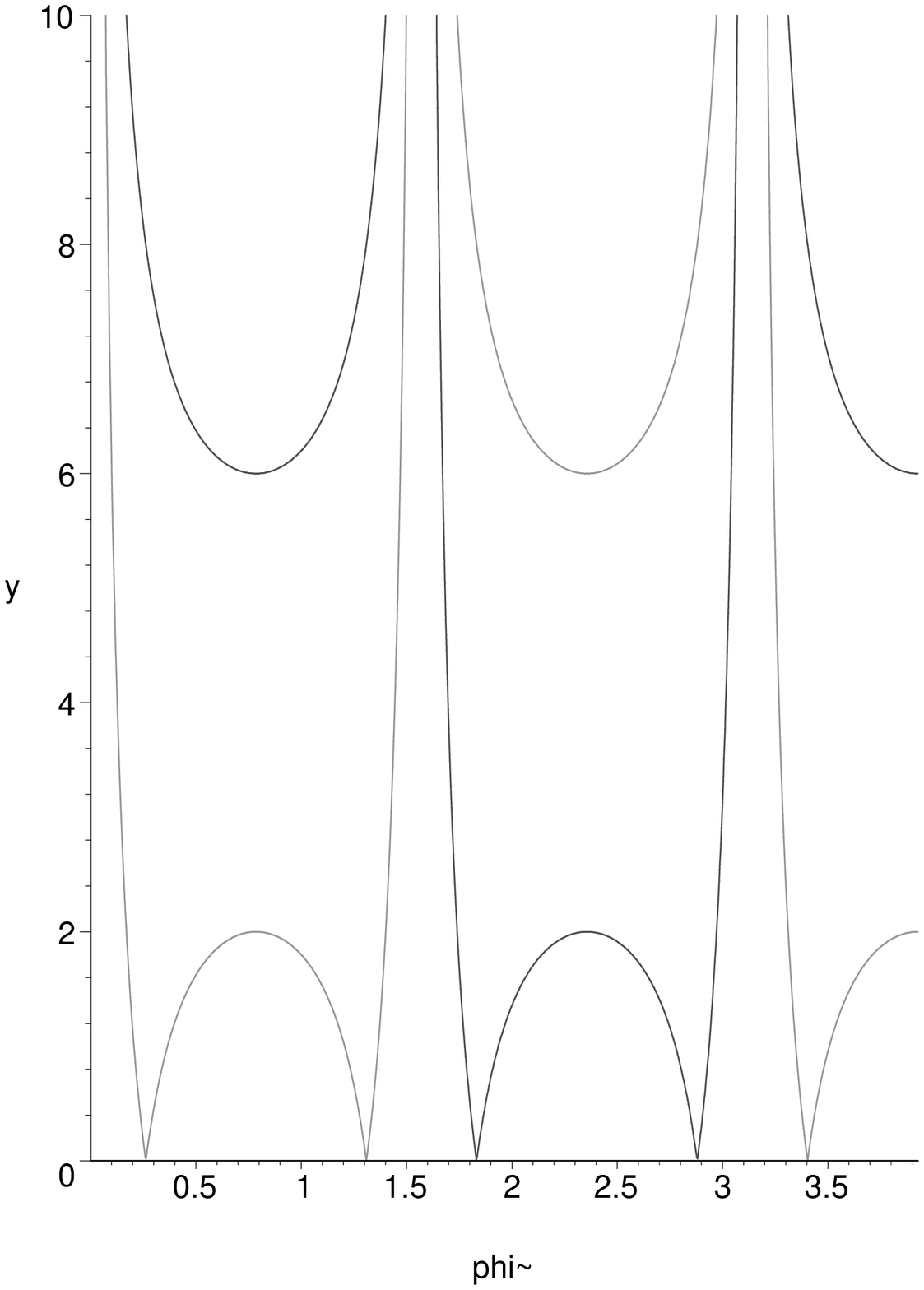,height=7truecm,width=8truecm}
\figskip
{\em Fig.~4: $M_1$ and $M_2$ as functions of $\varphi$, case $d=4>m_1=m_2=2$}
\end{center}
}

The singularities, where one of the masses goes to $\infty$,
correspond to $\rho \rightarrow \infty$.

On Figs.~5a, 5b and 6 are plotted respectively the ratio
$M_1/M_2$, $M_2/M_1$ and $|M_2-M_1|/(M_2+M_1)$ for $m=2$ and
$d=4$.

\figskip
\vbox{
\begin{center}
\epsfig{file=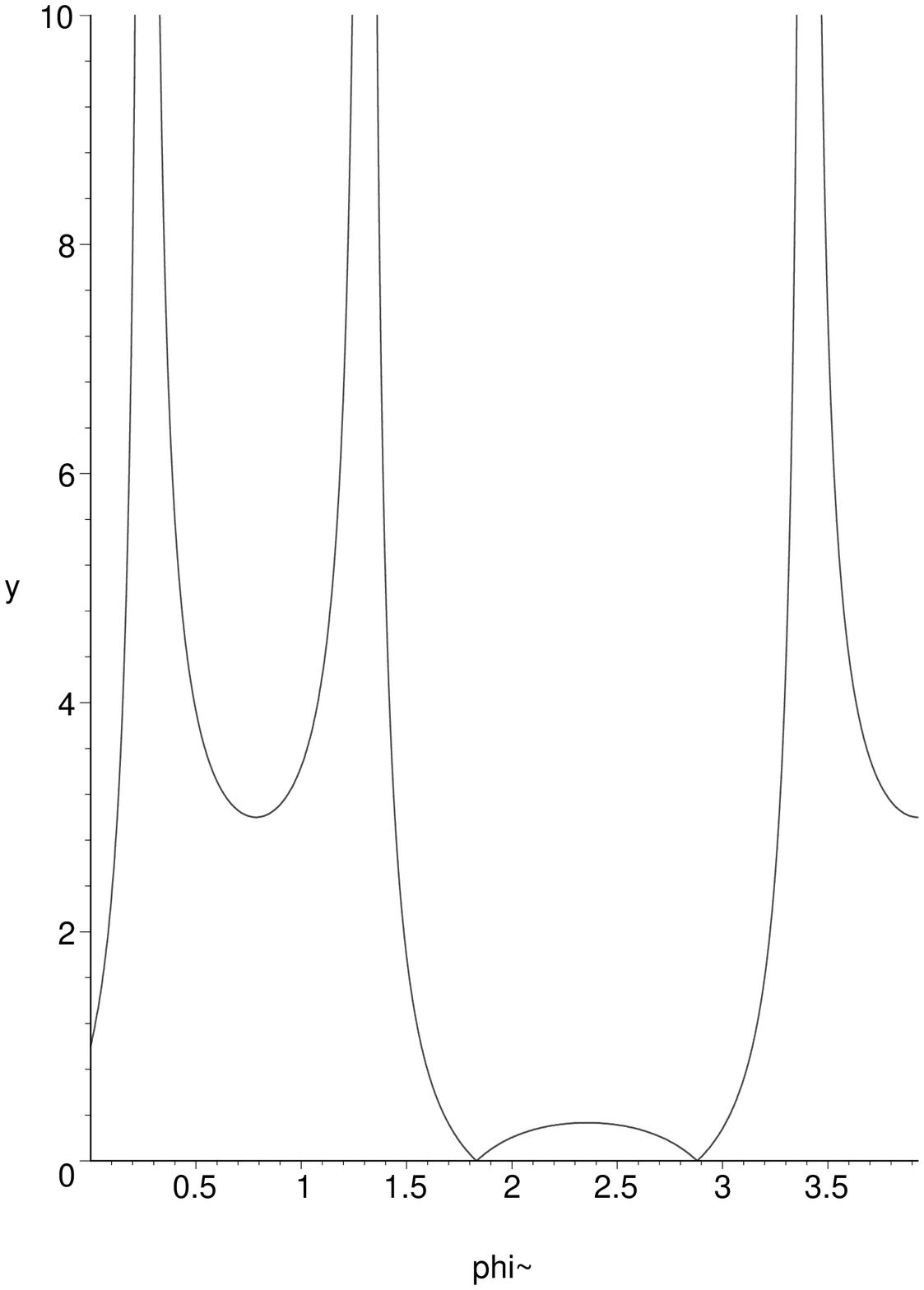,height=6truecm,width=8truecm} \figskip
{\em Fig.~5a: $M_1/M_2$ as a function of $\varphi$, case
$d=4>m_1=m_2=2$}
\end{center}
}
\figskip
\vbox{
\begin{center}
\epsfig{file=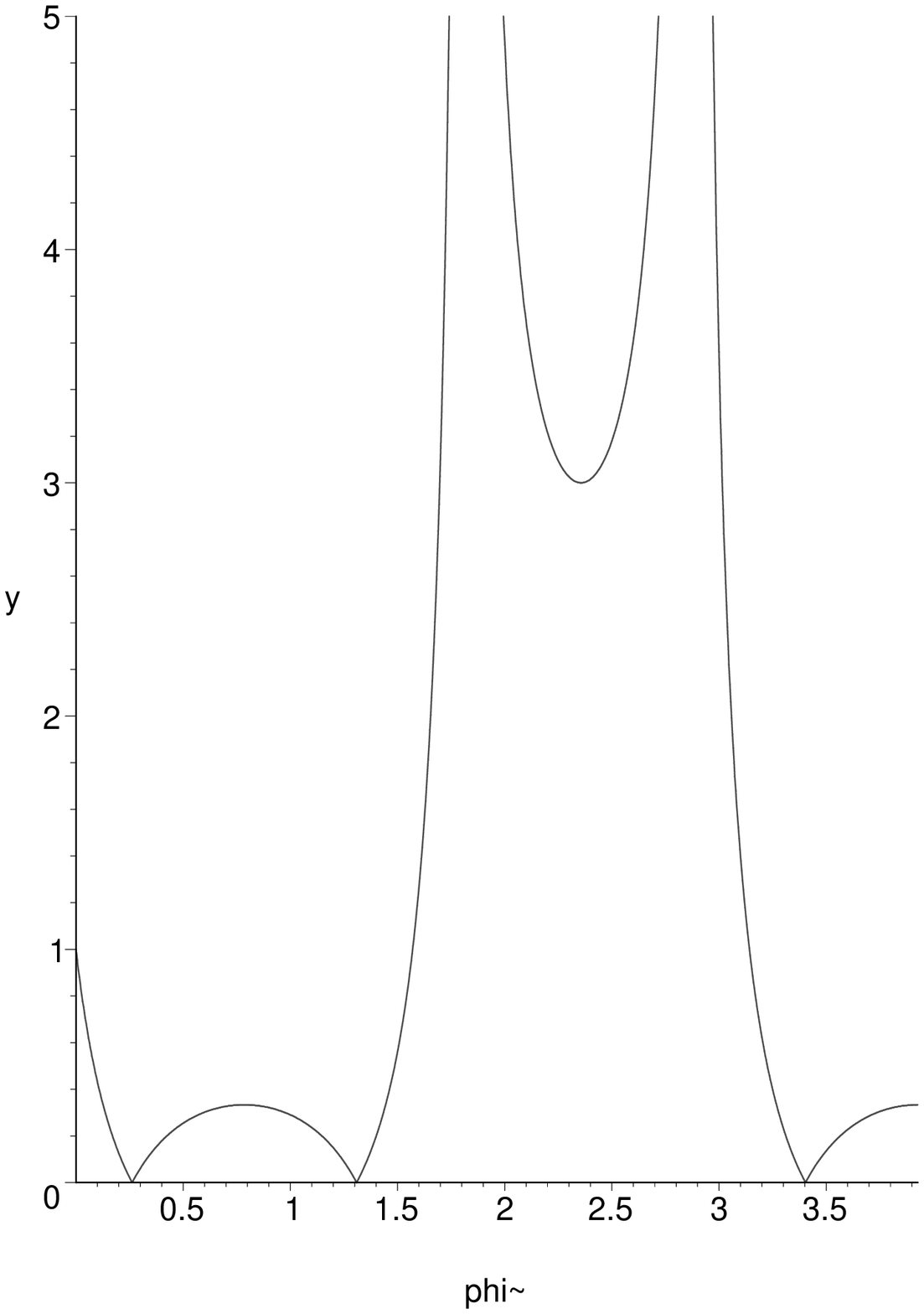,height=6truecm,width=8truecm} \figskip
{\em Fig.~5b: $M_2/M_1$ as a function of $\varphi$, case
$d=4>m_1=m_2=2$}
\end{center}
}
\figskip
\vbox{
\begin{center}
\epsfig{file=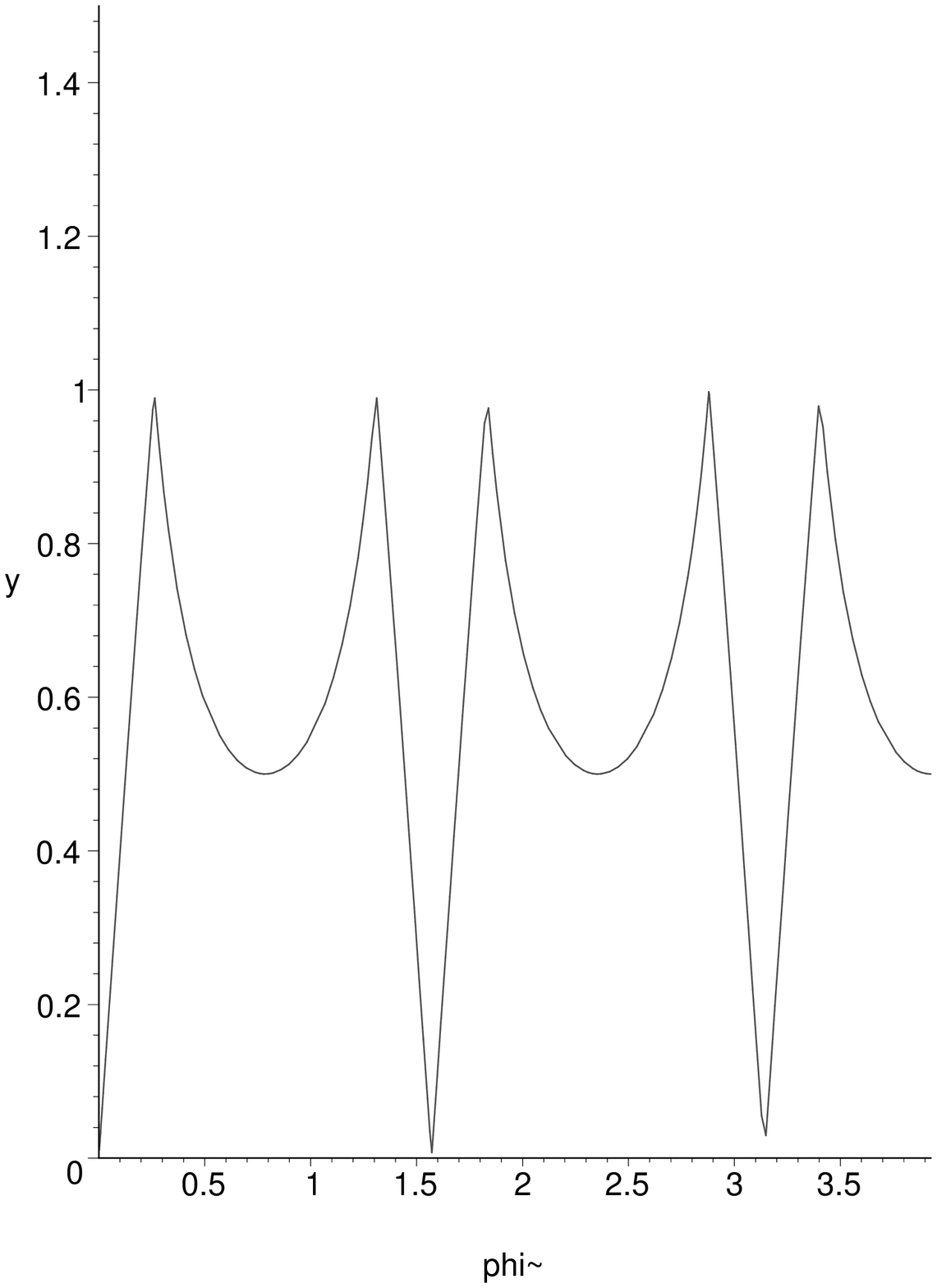,height=7truecm,width=8truecm} \figskip {\em
Fig.~6: $|M_2 - M_1|/(M_2+M_1)$ as a function of $\varphi$,
                            case $d=4>m_1=m_2=2$}
\end{center}
}
%
\subsubsection{Case $\mathbf{d<m}$}
\label{subsubsec:dinfm}

The spectrum is plotted on Fig.~7 for $m=4$ and $d=2$.
\figskip
\vbox{
\begin{center}
\epsfig{file=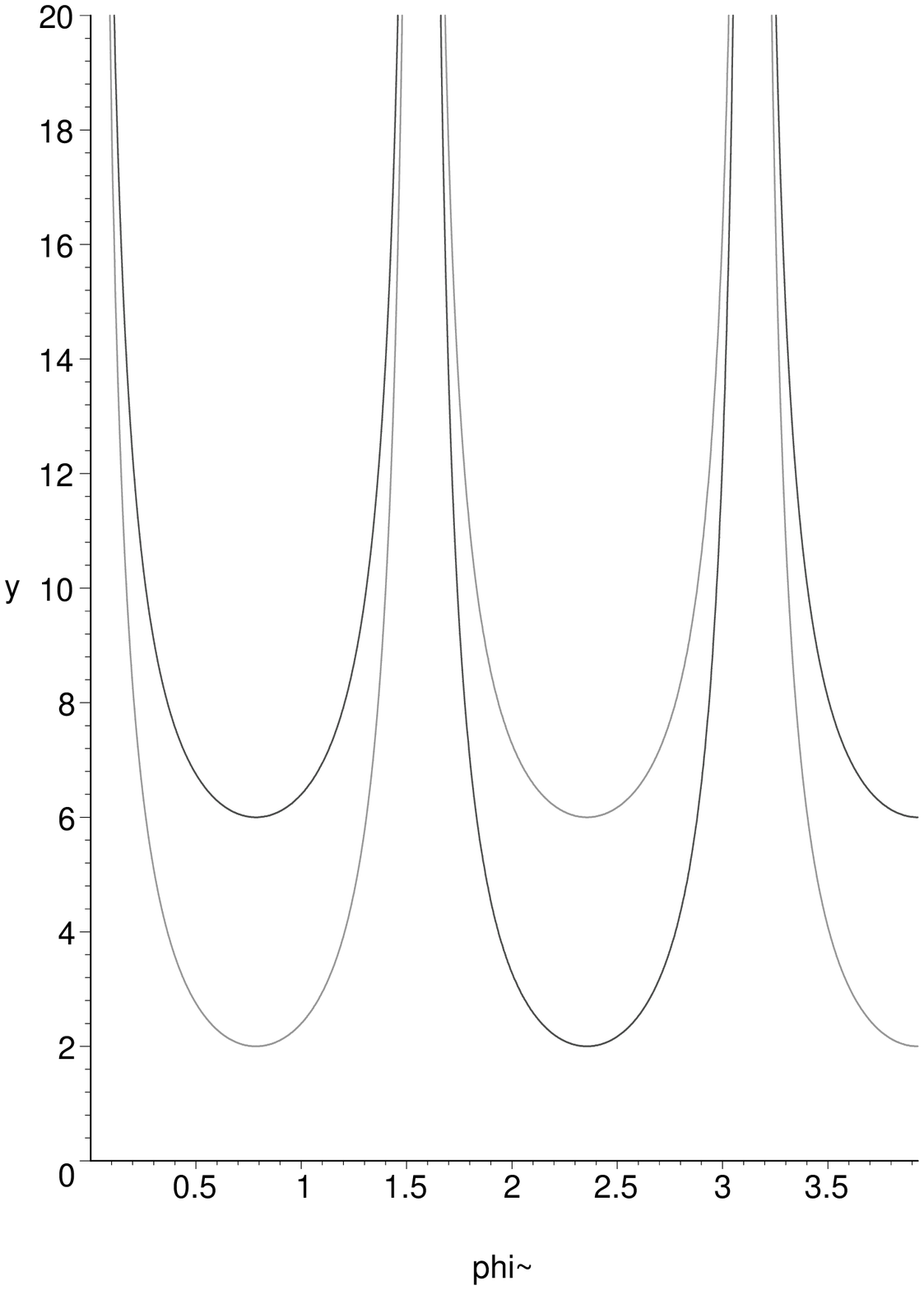,height=7truecm,width=8truecm}
\figskip
{\em Fig.~7: $M_1$ and $M_2$ as functions of $\varphi$, case $d=2<m_1=m_2=4$}
\end{center}
}
One has $M_1 + M_2 = \left\vert 2m/\sin(2\varphi)\right\vert$ and
$|M_2 - M_1| = 2d$: while the mass splitting is constant, the sum of the
masses reaches its minimum $2m$ for $\sin(2\varphi) = 1,
\rho=0$, in which case the mass matrix $\cal M$ is symmetric and the mass
eigenstates can also be taken as Majorana neutrinos.

None of the masses ever vanishes in this case, except when $m_1$
or $m_2$ vanishes.
The singularities where one of the masses goes to $\infty$
correspond to $\rho \rightarrow \infty$.

The ratio $M_1/M_2$ is plotted on Fig.~8.
\figskip
\vbox{
\begin{center}
\epsfig{file=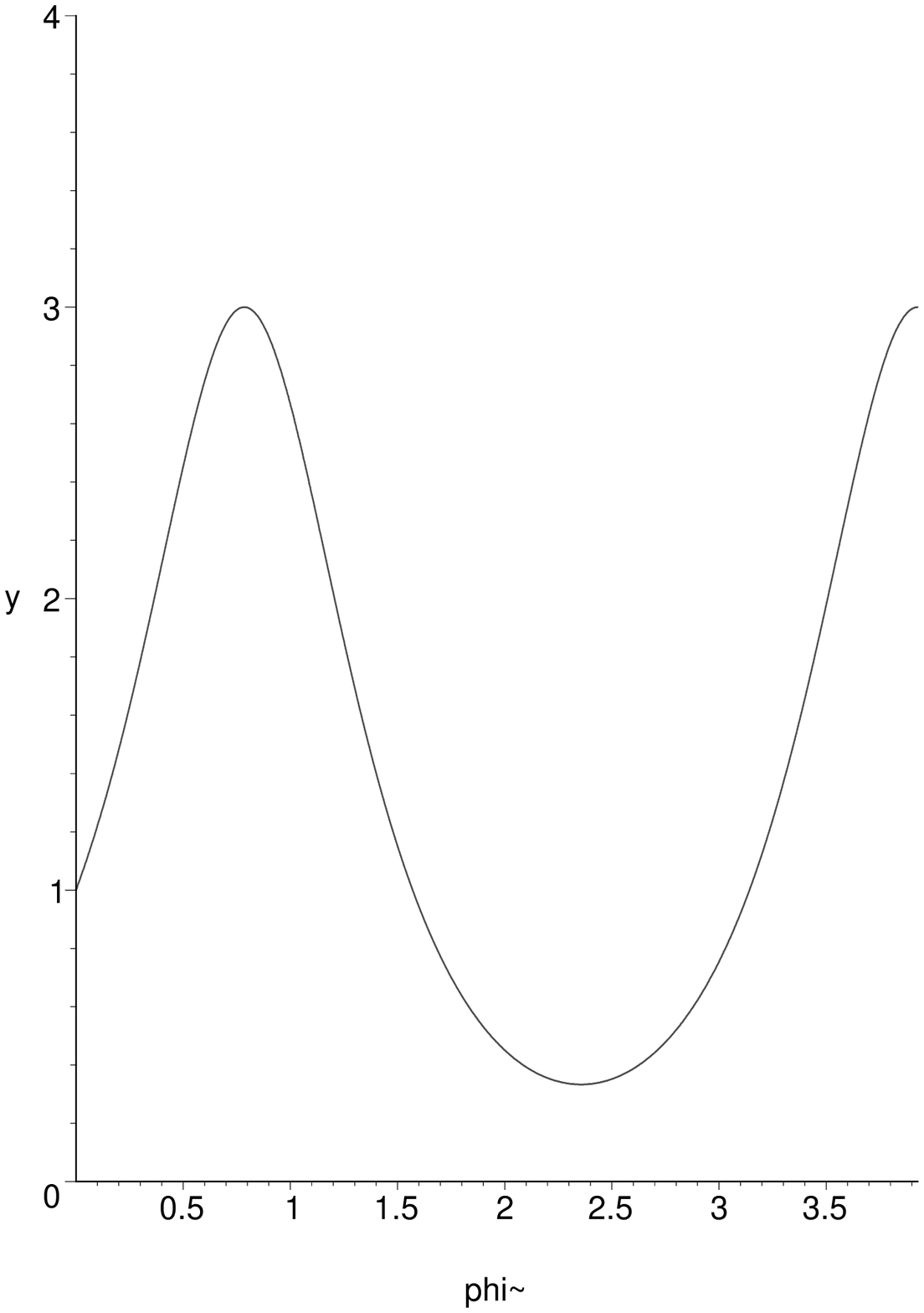,height=7truecm,width=8truecm} \figskip {\em
Fig.~8: $M_1/M_2$ as a function of $\varphi$, case
$d=2<m_1=m_2=4$}
\end{center}
}
Note the continuity of this curve through the singularities of Fig.~7.
%
\subsection{The case $\mathbf{m_2 \not = m_1,d=0}$}

 The corresponding mass matrix
\begin{equation}
{\mathfrak M} = \left( \begin{array}{rr}  m_1   &   \rho   \cr
                                     -\rho  &   m_2  \end{array}\right)
\end{equation}
is  diagonalised through the relation
\begin{equation}
B(\varphi)\; {\mathfrak M}\; A(\varphi) = diag(M_1,M_2),
\end{equation}
with
\begin{equation}
\tan(2\varphi) = \frac{2\rho}{m_1 + m_2}.
\label{eq:rho2}
\end{equation}
The two eigenmasses (which are easily made, as above, to be both
positive) are given by
\begin{eqnarray}
M_1 &=& \left\vert\frac{m_1 \cos^2\varphi + m_2 \sin^2\varphi}{\cos^2\varphi
- \sin^2\varphi}\right\vert, \cr
M_2 &=& \left\vert\frac{m_1 \sin^2\varphi + m_2 \cos^2\varphi}{\cos^2\varphi
- \sin^2\varphi}\right\vert,
\label{eq:mass2}
\end{eqnarray}
such that
\begin{equation}
M_1 + M_2 = \left\vert\frac{m_1 + m_2}{\cos^2\varphi -
\sin^2\varphi}\right\vert,  \quad
\vert M_2 - M_1 \vert = \vert m_2 - m_1 \vert.
\end{equation}
None of them ever vanishes since the determinant of $\mathfrak M$ cannot.

They are plotted on Fig.~9  for $m_1 =2$ and $m_2 = 4$.
\figskip
\vbox{
\begin{center}
\epsfig{file=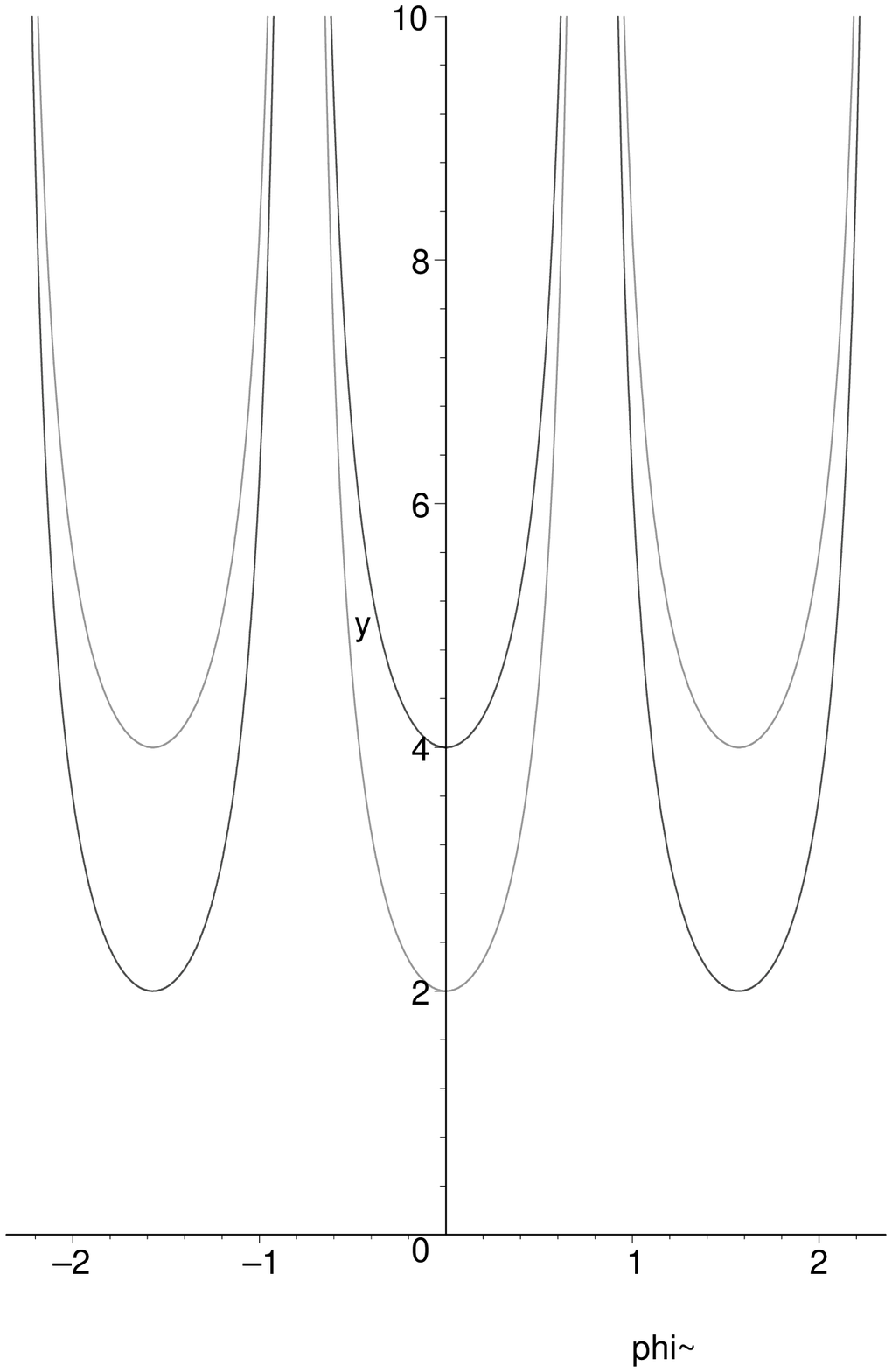,height=7truecm,width=8truecm} \figskip {\em
Fig.~9: $M_1$ and $M_2$ as functions of $\varphi$, case
$m_1=2$,$m_2=4$ and $d=0$}
\end{center}
}
The singularities where one of the masses goes to $\infty$ always
correspond to $\rho \rightarrow \infty$.

The ratio $M_2/M_1$ is plotted on Fig.~10.
Note, like for Fig.~8, the continuity of
this curve through the singularities of Fig.~9.
\figskip
\vbox{
\begin{center}
\epsfig{file=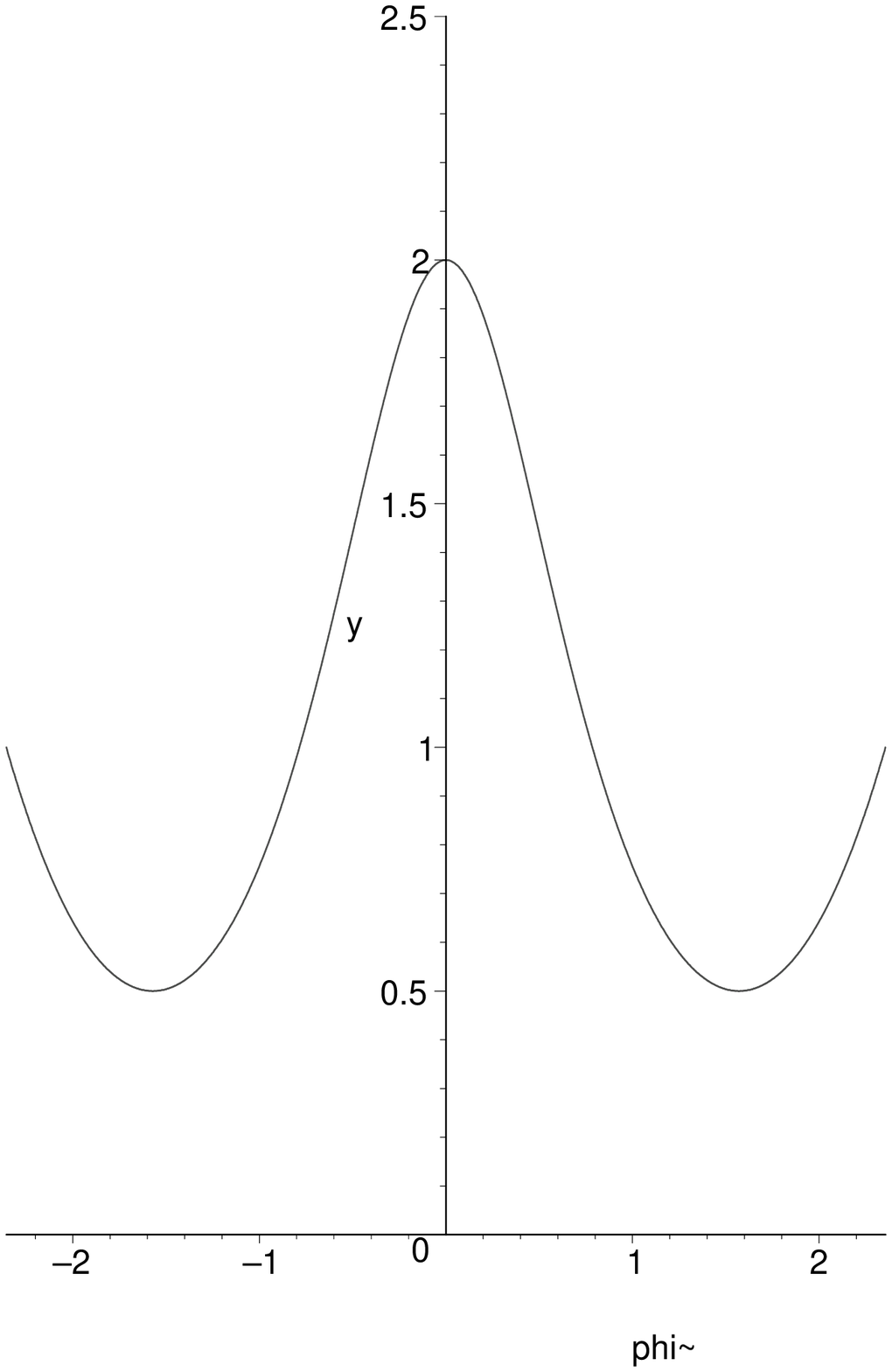,height=7truecm,width=8truecm} \figskip {\em
Fig.~10: $M_1/M_2$ as a function of $\varphi$, case $m_1=2$,
$m_2=4$ and $d=0$}
\end{center}
}
%
\subsection{The more general case $\mathbf{m_2 \not = m_1, d \not = 0}$}
\label{subsec:general}

The mass matrix (\ref{eq:Mgen}) can be diagonalised by the
bi-unitary transformation
\begin{equation}
B^\dagger(\omega) M A(\varphi) = diag(M_1,M_2),
\label{eq:dMg}
\end{equation}
such that
\begin{eqnarray}
\tan(\omega + \varphi) &=& \frac{2\rho}{m_1 + m_2};\cr
\tan(\omega - \varphi) &=& \frac{2d}{m_2-m_1}.
\label{eq:omegaphi}
\end{eqnarray}
%

The eigenmasses are \footnote{If one of the eigenmass turns out to
be negative, one multiplies $A$ and $B$ by the suitable diagonal
matrix $diag(1,i)$ or $diag(i,1)$.}
\begin{eqnarray}
M_1 &=& \left\vert\frac{m_1\cos\varphi\cos\omega + m_2\sin\varphi\sin\omega}
           {1- \cos^2\varphi - \cos^2\omega}\right\vert,\cr
M_2 &=& \left\vert\frac{m_1\sin\varphi\sin\omega + m_2\cos\varphi\cos\omega}
           {1- \cos^2\varphi - \cos^2\omega}\right\vert.
\label{eq:MM12}
\end{eqnarray}
On Fig.~11  is drawn the spectrum, $M_1$ and $M_2$ as functions of $\varphi$,
 for $m_1=2$, $m_2=4$ and $d=1$,
corresponding to $\tan(\omega - \varphi) =1$ or $ \omega = \varphi
+ \pi/4$.
\figskip
\vbox{
\begin{center}
\epsfig{file=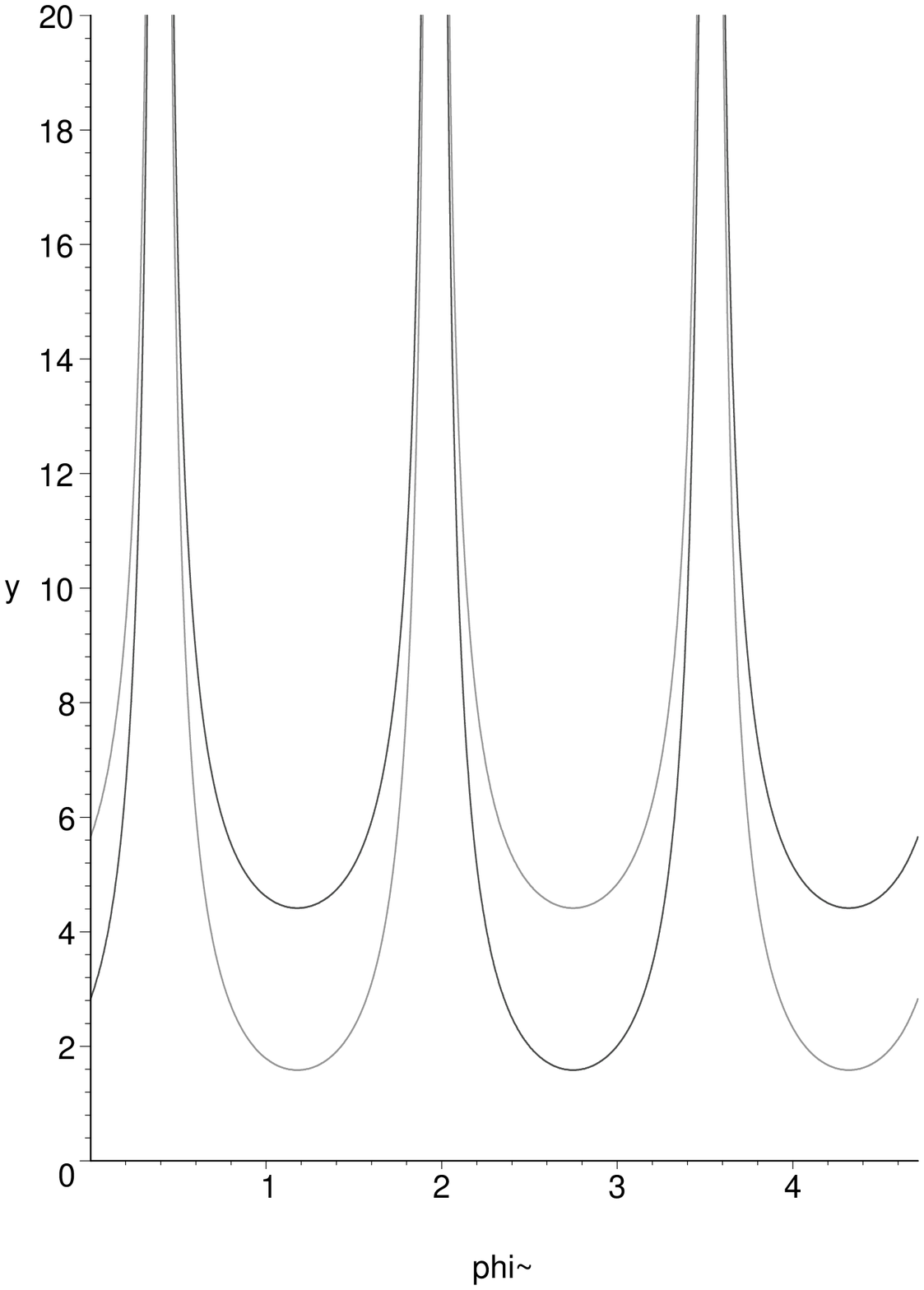,height=7truecm,width=8truecm}
\figskip
{\em Fig.~11: $M_1$ and $M_2$ as  functions of $\varphi$, for $m_1=2$,
$m_2=4$ and $\omega=\varphi+\pi/4$}
\end{center}
}
On Fig.~12 is drawn the spectrum for $m_1=2$, $m_2=4$ and $\omega= \varphi
+ 7\pi/15$.
\figskip
\vbox{
\begin{center}
\epsfig{file=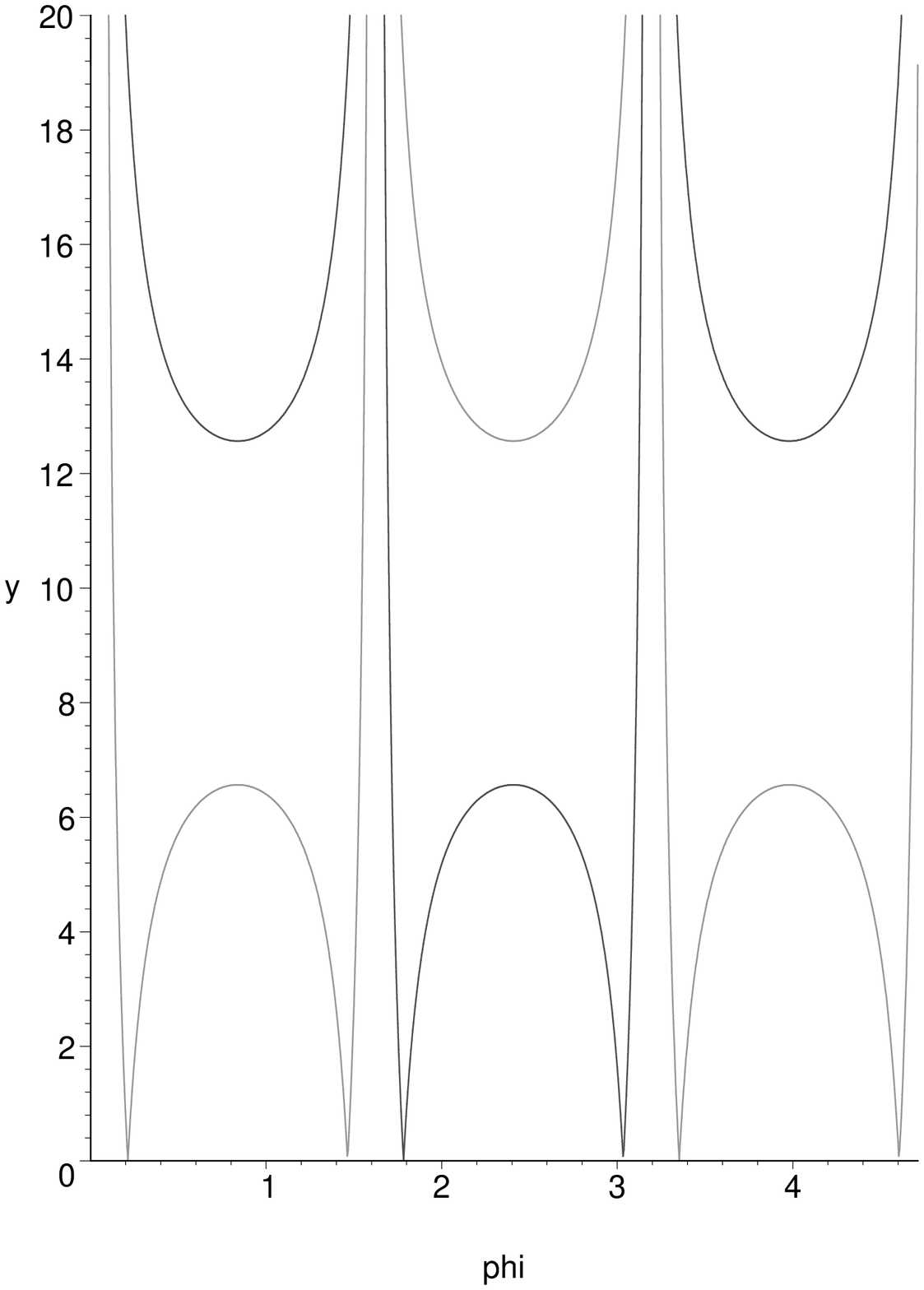,height=7truecm,width=8truecm}
\figskip
{\em Fig.~12: $M_1$ and $M_2$ as  functions of $\varphi$, for $m_1=2$,
$m_2=4$ and $\omega=\varphi+7\pi/15$}
\end{center}
}
The two sets of curves above correspond to fixing the value of $d$, which is
equivalent, by (\ref{eq:omegaphi}), to fixing the value of $\omega - \varphi$,
and to varying $\rho$.

The same types of spectra as in previous sections appear, only
shifted along the $\varphi$ axis.
For given $m_1$, $m_2$ and $d$ (corresponding to a ``standard'' mass
matrix), the hierarchy of masses depend now on $\rho$. On Fig.~13 is plotted
the ration $M_2/M_1$ as a function of $\varphi$ for the two sets of parameters
corresponding to the figures 10 and 11.
\figskip
\vbox{
\begin{center}
\epsfig{file=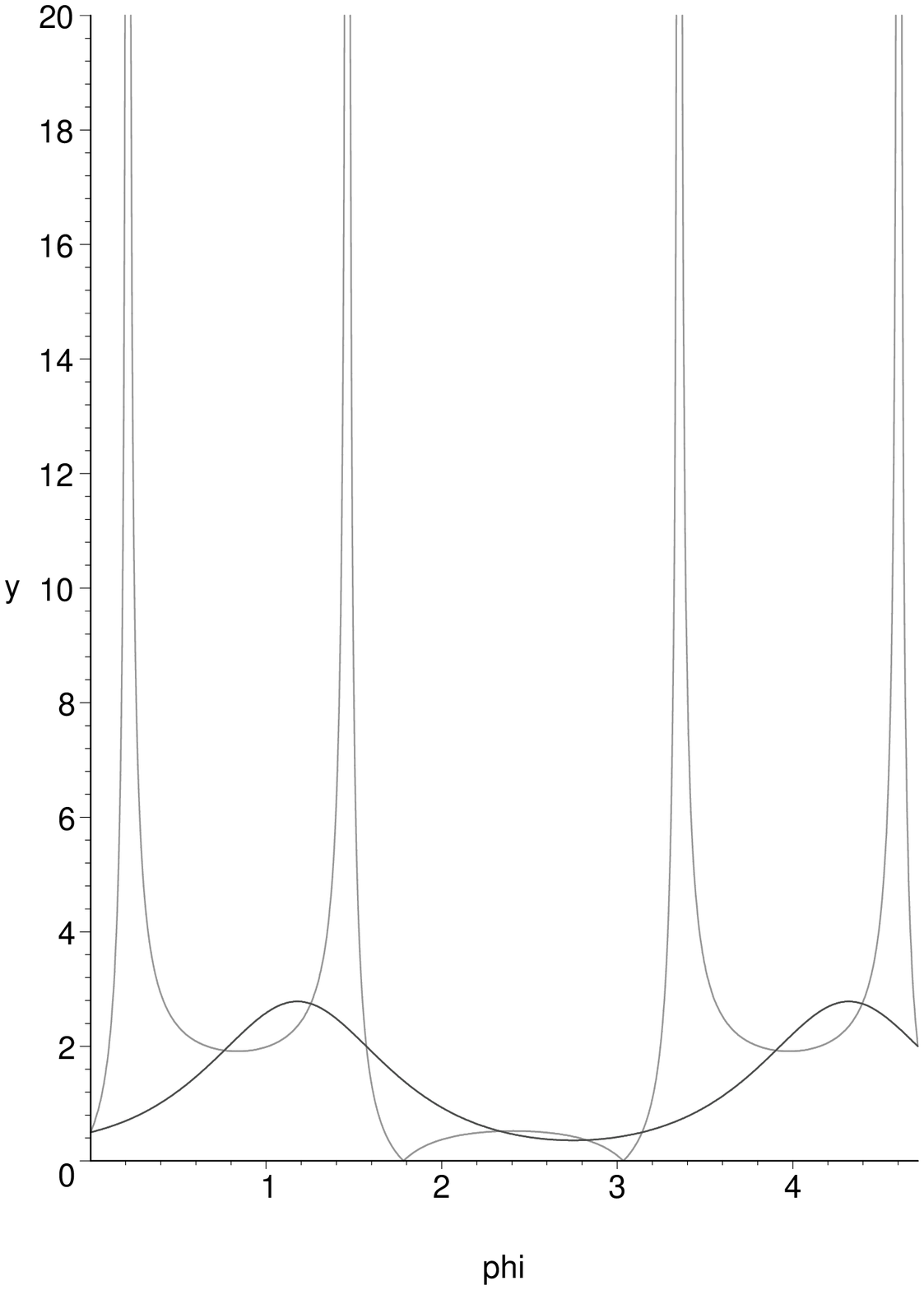,height=7truecm,width=8truecm}
\figskip
{\em Fig.~13: $M_2/M_1$ as  functions of $\varphi$, for the two sets of
parameters corresponding to Fig.~11 and Fig.~12}
\end{center}
}
On all examples of this section, the hierarchy of masses can
become very large ($\rightarrow \infty$) when the determinant of
the mass matrix vanishes. Tuning $\rho$ can consequently also have
the consequence that two among the four Weyl eigenstates decouple
from the theory at low enough energy. That they are not only
right-handed but form a left-right set distinguishes this
mechanism  from the usual see-saw.

\section{Weyl or Majorana?}
\label{section:WM}

Turning on the antisymmetric ``$\rho$'' term in $\cal M$ swaps
the mass eigenstates from Majorana to Weyl.

In the case of a symmetric mass matrix ($\rho = 0$), the two
points of view are, as shown below, equivalent. Let us deal with
the simple case of subsection \ref{subsec:degen} corresponding to
$m_2 = m_1 = m$.

The customary diagonalization method leading to Majorana mass
eigenstates uses the relation
\begin{equation}
\frac{1}{2} \left(\begin{array}{rr} 1 &  1 \cr
                                   -i &  i \end{array}\right)
            \left(\begin{array}{rr} m &  d \cr
                                    d &  m \end{array}\right)
            \left(\begin{array}{rr} 1 & -i \cr
                                    1 &  i \end{array}\right) =
            \left(\begin{array}{cc} m+d & 0 \cr
                                     0  & d-m \end{array}\right),
\label{eq:dia1}
\end{equation}
which corresponds, in (\ref{eq:V0}), to
\begin{equation}
V^0 = \frac{1}{\sqrt{2}}\left(\begin{array}{rr} 1  &  -i  \cr
                                                1  &   i \end{array}\right)
=\left(\begin{array}{rr} \cos\pi/4  &  -\sin\pi/4  \cr
                         \sin\pi/4  &   \cos\pi/4 \end{array}\right)
\left(\begin{array}{rr} 1  &  0  \cr
                        0  &  i \end{array}\right).
\label{eq:di1}
\end{equation}
The diagonal matrix $diag(1,i)$ in (\ref{eq:di1}) ensures the
positivity of both eigenmasses.

The diagonalization method advocated for in this work, which
leads to left- and right- Weyl mass eigenstates, uses instead the
relation (we consider the case $d>m$, such that $d-m$ is a
positive eigenmass)
\begin{equation}
\frac{1}{2} \left(\begin{array}{rr} 1 &  1 \cr
                                    1 &  -1 \end{array}\right)
            \left(\begin{array}{rr} m &  d \cr
                                    d &  m \end{array}\right)
            \left(\begin{array}{rr} 1 & -1 \cr
                                    1 &  1 \end{array}\right) =
            \left(\begin{array}{cc} m+d & 0 \cr
                                     0  & d-m \end{array}\right),
\label{eq:diag2}
\end{equation}
and corresponds, in (\ref{eq:bidiag}), to
\begin{equation}
V = \left(\begin{array}{rr} 1  &  -1  \cr
                            1  &   1 \end{array}\right),\quad
U = \left(\begin{array}{rr} 1  &  1   \cr
                            1  &  -1 \end{array}\right).
\end{equation}
The Majorana mass eigenstates obtained from the customary diagonalization
are given by
\begin{eqnarray}
N_1^{Maj} &=& \frac{1}{\sqrt{2}}\left(( n_{1L} + n_{2L} ) +( n_{1L} + n_{2L}
)^c\right), \cr
N_2^{Maj} &=& \frac{1}{\sqrt{2}}\left((i\,n_{1L} - i\, n_{2L}) +(i\,n_{1L}
- i\, n_{2L})^c\right),
\label{eq:majeigen}
\end{eqnarray}
while the Weyl mass eigenstates obtained in subsection \ref{subsec:degen}
are
\begin{eqnarray}
N_{1L}^{Weyl} &=& \frac{1}{\sqrt{2}}(n_{1L} + n_{2L})
                                               = (N_1^{Maj})_L,\cr
N_{2L}^{Weyl} &=& \frac{1}{\sqrt{2}}(-n_{1L} + n_{2L})
                                               = (i\, N_2^{Maj})_L,\cr
N_{1R}^{Weyl} &=& \frac{1}{\sqrt{2}}(n_{1L} + n_{2L})^c
                                               = (N_1^{Maj})_R,\cr
N_{2R}^{Weyl} &=& \frac{1}{\sqrt{2}}(n_{1L} - n_{2L})^c
                                               = (-i\, N_2^{Maj})_R.
\label{eq:direigen}
\end{eqnarray}
Since the latter satisfy $N_{1R}^{Dir} = (N_{1L}^{Dir})^c$ and
$N_{2R}^{Dir} = -(N_{2L}^{Dir})^c$, the two formulations (Weyl and
Majorana) become equivalent for $\rho = 0$.
%
\section{Mixing angles}
\label{section:angles}

One of the results of this work is that, unlike usually
considered, small mixing angles are not {\em a priori} attached to
large hierarchies of masses, neither, as a consequence, large
mixing angles to a near degeneracy or ``inverted'' hierarchies.
Let us be more explicit.

In the customary framework of a symmetric mass matrix (see Fig.~2b
for example), the hierarchy $M_1/M_1$ starts by increasing (up to
$\infty$) with the mixing angle, then decreases down to $1$, gets
inverted, goes  down to $0$, to finally (still inverted) increase
again up to $1/2$.

Consider next the example of Fig.~5a. The mass hierarchy $M_1/M_2$
is minimal (close to $1$) for $\varphi$ small, increases to
infinity for $\sin(2\varphi)= m/d$ and then decreases when
$\varphi$ increases  up to the ``maximal mixing'' value $\varphi=
\pi/4$. After this, it increases again to $\infty$, goes down to
$1$, gets inverted down to $0$ to finally slightly increase again.
 Note that the  hierarchy goes to $1$ when the
two masses become very large, for $\varphi \rightarrow 0\ or \
\pi/2$.

The next example is that of Fig.~8, where a large mixing angle
($\pi/4$) is  associated with the largest hierarchy $M_1/M_2 = 3$.

Accordingly, the ``Large Mixing Angle'' solution for leptons
\cite{LMA} can also go along with large hierarchies among neutrino
masses.

In the last example,  Fig.~10,  in agreement with the common
prejudice, the largest hierarchies ($2$ or $1/2$ in this case) occur for
a small mixing angle $\varphi = k\pi/2$, and the smallest hierarchy
($1$), occurs for maximal mixing $\varphi = \pi/4 + k\pi/2$.

In the more general case of subsection \ref{subsec:general}, the variety of
the shifts than the spectrum can undergo along the mixing angle axis can
produce {\em a priori} all kinds of variations of the mass hierarchy as
a function of the mixing angle.

The relation between the two appears consequently as dependent of
the structure of the mass matrix, and, specially, of the presence or not of
the antisymmetric ``$\rho$'' term.

\section{Symmetries}
\label{section:symmetries}

In general (when $\rho \not= 0$), $CP$ is indirectly violated;
the ``left'' and ``right'' mass eigenstates (Weyl spinors) are no longer
$CP$ eigenstates, and can be written as linear combinations
of the Majorana ($CP$) eigenstates which correspond to $\rho=0$,
projected on a given helicity.
Apart from the spin degrees of freedom, the similarity with the neutral
kaon system clearly appears.

Like in \cite{Machet1}, a $U(1)$ group of transformation can be associated
with $CP$.
The corresponding phase is the angle $\theta$ defined in (\ref{eq:V0}).
Though the eigenmasses depend on the value of the antidiagonal symmetric
$d$ term and thus on $\theta$, whatever be the latter,
the eigenvectors of the mass matrix $M_0$ given in
(\ref{eq:M0}), where no antidiagonal antisymmetric $\rho$ term is present,
are always Majorana neutrinos, independent of $\theta$;
hence they are also $CP$ eigenstates.
Any choice for $\theta$ breaks this $U(1)$ symmetry but,
since the eigenstates stay the same $CP$ eigenstates, it does not break $CP$
invariance; it only selects a specific mass pattern.

$\theta$ becomes the ``$CP$'' violating phase only when the $\rho$ term is
turned on, whatever its value (even when it goes to $0$); the breaking
of $U(1)$ and $CP$ are consequently connected.\l
Whether this breaking is spontaneous or explicit is an ambiguous
matter. For a non-vanishing commutator, $U(1)$ and $CP$ are
simultaneously explicitly broken when $\rho$ is turned on.
However, for independent generations, it amounts to introducing a
vanishing perturbation;  one is inclined in this case to consider
that both symmetries become spontaneously broken
\footnote{Different ``domains'' corresponding to different values of
$\theta$ can exist in disconnected space-time regions; the $U(1)$ and $CP$
symmetries can be expected to be restored only by  averaging over the whole
sets of domains.}.

The problem is twofold since one needs not only to fix
$\rho$, but also to know whether fermions belonging to different generations
truly anticommute.

What fixes $\rho$?  The example of the neutral kaon system
taught us \cite{Machet1} that, if they are taken as composite
and if their constituents undergo another interaction which is
misaligned with flavor, the relevant commutator is expected not to
vanish; the other interactions are, for the kaon system, electroweak
interactions
\footnote{The term in the Lagrangian proportional to the
 commutator of $K^0$ and $\overline{K^0}$ was connected to
the (integral of the) number of $d$ quarks minus the number of
$s$ quarks, which is not conserved by electroweak
interactions and ``penguin''-type diagrams.
This, however, did not tell us what would fix the value of $\rho$.}.

In the present case, the situation is more difficult, since one
is reluctant to introduce for leptons another level of constituents.

A reasonable  attitude is to consider  that an ambiguity in the mass spectrum
cannot exist and to interpret the paradox found above  as a theoretical
signal that nature has probably not made generations truly independent and
anticommuting.

To go further in this direction, a conventional idea is to introduce
additional, eventually gauged, horizontal interactions.
They are likely to set up a new fundamental length scale,
and it may be not unrealistic to think that it could be connected to $\rho$
\footnote{
But are they really necessary since, when neutrinos are massive,
flavor-changing (like $\mu-e$) transitions are induced through
electroweak loops?}.

Then, the question arises of how the anticommutator of two
fermions belonging to different generations can be determined. In
terms of which fields? Is the commutator to be considered itself
as a new (composite) field eventually coupled, as we did, to
fermions? Can it be considered as independent or should
constraints be introduced?

The situation is evidently  far from clear and shedding light on it
is evidently beyond the scope of this work.

We also mentioned in \cite{Machet1}, in the case of composite
Higgs-like doublets, that, for one of them, which includes the
scalar flavor singlet and its three pseudoscalar partners, the
ambiguity remains and its mass stays undetermined; this doublet
commutes indeed with all other doublets of the same type. The
eventual occurrence of the same phenomenon in the fermionic case,
and the presence of a robust ambiguity, in the case of truly independent and
anticommuting generations of fermions, should consequently not be
overlooked.

\section{Conclusion}
\label{section:conclusion}

Following  a similar line of argumentation as we did for the neutral
kaon system, we have shown that the existence of truly independent,
anticommuting, generations of fermions leads to an ambiguity
in the mass spectrum of neutrinos.

The introduction of the parameter $\rho$ with the dimension
of $[mass]$, closely related to the replication of families,
sets a strong correlation  between mixing angles, $CP$
violation,  and the hierarchy of masses.

The presence of $\rho$ brings a new point of view on the presence
of large mixing angles for leptons \cite{LMA} which are no longer
preferentially associated with the smallest hierarchies
\cite{MachetPetcov}.

If more generations are present, several arbitrary $\rho$-like parameters
can be introduced; the variation of any among them will modify the mass
pattern of neutrinos.

The main problem is that, for independent generations, $\rho$
cannot play any role since it corresponds to a vanishing
interaction term in the Lagrangian.

A first possible attitude would be to  prove that introducing such
terms is forbidden in field theory.

Otherwise, breaking this indetermination most likely requires
that the generations be not independent, hence  not anticommuting;
the simplest assumption is, in analogy with the neutral kaon system,
that subconstituents of fermions undergo a new type of interaction which is
misaligned with electroweak interactions.
However,  introducing a new level of constituents for leptons is unwelcome,
unless several other converging compelling constraints urge to do it.

%
\vskip 1cm
\begin{em}
\underline {Acknowledgments}: It is a great pleasure to thank
the group of L. Okun, V. Novikov, M. Vysotsky \ldots at
ITEP, where a large part of this work was completed, for the warm
hospitality extended to me.
\end{em}
\vskip .5cm
\newpage\null
\listoffigures
\bigskip
\begin{em}
Fig.~1: $M_1$ and $M_2$ as functions of $\theta$, for $m_1=2$,
                 $m_2=4$ and $\rho=0$;\l
Fig.~2~a,b: $M_1/M_2$ and $M_2/M_1$ as functions of $\theta$
           for $m_1=2$, $m_2=4$ and $\rho=0$;\l
Fig.~3:  propagating the Weyl neutrino $N_{1L}$;\l
Fig.~4:  $M_1$ and $M_2$ as functions of $\varphi$, case $d=4 >m_1=m_2=2$;\l
Figs.~5~a,b:  $M_1/M_2$ and $M_2/M_1$ as  functions of $\varphi$,
           case $d=4>m_1=m_2=2$;\l
Fig.~6:  $\vert M_2 - M_1\vert/(M_2+M_1)$ as a
        function of $\varphi$, case $d=4>m_1=m_2=2$;\l
Fig.~7:  $M_1$ and $M_2$ as functions of $\varphi$, case $d=2<m_1=m_2=4$;\l
Fig.~8:  $M_1/M_2$ as a function of $\varphi$, case $d=2<m_1=m_2=4$;\l
Fig.~9:  $M_1$ and $M_2$ as functions of $\varphi$, case $m_1=2$, $m_2=4$ and
                         $d=0$;\l
Fig.~10:  $M_1/M_2$ as a function of $\varphi$, case $m_1=2$, $m_2=4$ and
                $d=0$;\l
Fig.~11: $M_1$ and $M_2$ as functions of $\varphi$, for $m_1=2$,
              $m_2=4$ and $\omega=\varphi+\pi/4$;\l
Fig.~12: $M_1$ and $M_2$ as functions of $\varphi$, for $m_1=2$,
              $m_2=4$ and $\omega=\varphi+7\pi/15$;\l
Fig.~13: $M_2/M_1$ as  functions of $\varphi$, for the two sets of
                 parameters corresponding to Figs.~10,11.
\end{em}
%
%
\newpage\null
\begin{em}

\end{em}

\end{document}